\documentclass[12pt]{article}
\usepackage{epsf}
\def\beq{\begin{equation}}
\def\eeq{\end{equation}}

\begin{document}
\begin{titlepage}
\begin{center}
{\Large \bf William I. Fine Theoretical Physics Institute \\
University of Minnesota \\}
\end{center}
\vspace{0.2in}
\begin{flushright}
FTPI-MINN-07/14 \\
UMN-TH-2601/07 \\
April 2007 \\
\end{flushright}
\vspace{0.3in}
\begin{center}
{\Large \bf Isospin properties of the $X$ state near the $D {\bar D}^{*}$
threshold
\\}
\vspace{0.2in}
{\bf M.B. Voloshin  \\ }
William I. Fine Theoretical Physics Institute, University of
Minnesota,\\ Minneapolis, MN 55455 \\
and \\
Institute of Theoretical and Experimental Physics, Moscow, 117218
\\[0.2in]
\end{center}

\begin{abstract}

The $D {\bar D}^*$ scattering amplitude and the production of the final states
$\pi^+ \pi^- J/\psi$ and $\pi^+ \pi^- \pi^0 J/\psi$ near the $D^0 {\bar D}^{*0}$
threshold are discussed following the recent suggestion that the observed peaks
$X(3872)$ and $X(3875)$ in the decays $B \to X \, K$ are due to a virtual state
$X$ in the $D^0 {\bar D}^{*0}$ channel. The strong interaction is treated using
the small interaction radius approximation. It is shown that the mass difference
between the charged and neutral $D^{(*)}$ mesons results in a distinctive
behavior of the relevant isotopic amplitudes. In particular, the shape of the
peak in the $\pi^+ \pi^- J/\psi$ channel should be significantly narrower than
in the $\pi^+ \pi^- \pi^0 J/\psi$ channel, which property can be used for an
experimental test of the virtual state hypothesis.

\end{abstract}

\end{titlepage}

The narrow peak $X(3872)$\,\cite{belle1,cdf,d0,babar1} currently commands a
great interest due to the suggested possibility that it is dominantly a
molecular state\,\cite{ess,nat,mv1} of charmed mesons $(D^0 {\bar D}^{*0} +
D^{*0} {\bar D}^0)$, probably of the type discussed in the literature long time
ago\,\cite{ov, drgg}. Such interpretation, as opposed to considering $X(3872)$
as a regular charmonium resonance, is essentially based on two remarkable
observations: the extreme proximity of the mass of $X$ to the $D^0 {\bar
D}^{*0}$ threshold and the apparently strong violation of the isotopic symmetry
indicated by the co-existence of the decays $X(3872) \to \pi^+ \pi^- J/\psi$ and
$X(3872) \to \pi^+ \pi^- \pi^0 J/\psi$\,\cite{belle2,belle3}. The measurement of
the position of the $X$ relative to the $D^0 {\bar D}^{*0}$ threshold has been
recently improved by the CLEO result\,\cite{cleo} for the $D^0$ mass, which
places the $D^0 {\bar D}^{*0}$ threshold at $3871.81 \pm 0.36\,$MeV, and
corresponds to $M(X)-M(D^0 {\bar D}^{*0})=-0.6 \pm 0.6\,$MeV. Due to the
exceptional closeness of the peak to the meson-pair threshold one can expect
that the mass difference between the pairs of charged and neutral mesons,
$\Delta = M(D^+ D^{*-}) - M(D^{0} {\bar D}^{*0}) \approx 8.1\,$MeV, should
naturally give rise to a significant isospin breaking in the properties of the
state $X$.

Furthermore, recent experimental study of the $B$ meson decays $B \to D^0 {\bar
D}^0 \pi^0 \, K$\,\cite{belledd,babardd} and $B \to D^0 {\bar D}^0 \gamma \,
K$\,\cite{babardd} revealed that the invariant mass recoiling against the Kaon
displays a significant enhancement with a maximum at approximately 3875\,MeV,
which is only about 3\,MeV above the $D^0 {\bar D}^{*0}$ threshold. The observed
events can all be in fact attributed to the process $B \to (D^0 {\bar D}^{*0} +
{\bar D} D^{*0}) \, K$ since no distinction between the $D^{*0}$ mesons and
their decay products was done. Moreover, the yield of the heavy meson pairs
within the above-threshold peak is about ten times larger that that of the
$\pi^+ \pi^- J/\psi$ and $\pi^+ \pi^- \pi^0 J/\psi$ channels at the peak of
$X(3872)$. It has been most recently argued\,\cite{hkkn} that a very plausible
explanation of the observed enhancement of the $D^0 {\bar D}^{*0}$ production
combined with the smaller observed $X(3872)$  peak in the $\pi^+ \pi^- J/\psi$
channel is that both these phenomena are due to a virtual state\,\cite{bugg,yuk}
in the $D^0 {\bar D}^{*0}$ channel. In this picture the observed peak in the
$\pi^+ \pi^- J/\psi$ and $\pi^+ \pi^- \pi^0 J/\psi$ mass spectra is in fact a
cusp with a sharp maximum at the $D^0 {\bar D}^{*0}$ threshold.

In this paper such possible virtual state near the $D {\bar D}^*$ threshold is
discussed within the approximation of small interaction radius, and a
consideration is given to the isospin properties of the scattering and
production amplitudes in this energy region. This approach is similar to the
previously pursued\,\cite{bk} `universal scattering length' approximation, and
differs in including the effect of the nearby threshold for charged mesons $D^+
D^{*-}$. An interesting energy-dependent behavior of the isotopic properties
arises from the mere fact of the mass splitting $\Delta$ between the two
isospin-related and coupled $D {\bar D}^*$ channels. In particular it will be
argued that the expected pattern of the isospin breaking is consistent with the
observed relative yield of $\pi^+ \pi^- J/\psi$ and $\pi^+ \pi^- \pi^0 J/\psi$
at the peak. Moreover, it will be shown that the production amplitude for the
$I=1$ state $\pi^+ \pi^- J/\psi$ in the considered approximation necessarily has
a zero between the $D^0 {\bar D}^{*0}$ and $D^+ {D}^{*-}$ thresholds, thus
reducing the apparent width of the cusp and putting it in line with the
experimental limit\,\cite{belle1} $\Gamma < 2.3\,$MeV on the width of the peak
in this particular channel.

It is assumed throughout this paper that the quantum numbers associated with the
discussed peaks $X(3872)$ and $X(3875)$ are $J^{PC}=1^{++}$, corresponding to
the $S$ wave motion in the $C$ even state $D^0 {\bar D}^{*0} + {\bar D} D^{*0}$
and the coupled channel with charged mesons $D^+ D^{*-}+D^- D^{*+}$. These two
channels are referred, for brevity, as respectively $n$ and $c$. Considering the
nonrelativistic dynamics of the mesons it is convenient to place the origin of
the c.m. energy $E$ at the threshold in the $n$ channel: $E=M(D {\bar D}^*) -
M(D^0)- M(D^{*0})$. The energy range of interest for the present discussion is
from few MeV below the $n$ threshold and up to the $c$ threshold, i.e. up to $E
\approx \Delta$. In this range the  scale of the c.m. momentum (real and
virtual) in either channel is set by $\sqrt{2 \mu \Delta} \approx 127\,$MeV,
where $\mu \approx 970\,$MeV is the reduced mass for the meson pair. One can
apply in this region of soft momenta the standard picture of the strong-
interaction scattering (see e.g. in the textbook \cite{ll}), where the strong
interaction is localized at distances $r < r_0$ such that $r_0 \, \sqrt{2 \mu
\Delta}$ can be considered as a small parameter. Some well known points of this
time-tested approach are repeated here in order to adapt the same treatment to
the situation with two closely spaced thresholds in isotopically related
channels. Outside the region of the strong interaction the motion the $n$ and
$c$ channel is described by the free $S$-wave radial wave function. Considering
for definiteness an energy value between the two thresholds, $0 < E < \Delta$,
one can write the corresponding wave functions (up to an overall normalization
constant) as \beq \chi_n(r)=\sin (k_n r +\delta), ~~~~~\chi_c(r) = \xi \, \exp
(-\kappa_c r)~, \label{chinc} \eeq where $k_n = \sqrt{2 \mu E}$ and
$\kappa_c=\sqrt{2 \mu (\Delta - E)}$, $\delta$ is the elastic\footnote{In this
consideration the small inelasticity due to the $\pi^+ \pi^- J/\psi$ and $\pi^+
\pi^- \pi^0 J/\psi$ channels is neglected and will be included later. Also the
small width of the $D^*$ mesons is entirely neglected throughout this paper.}
scattering phase in the $n$ channel and the constant $\xi$, generally energy-
dependent, describes the relative normalization and phase of the wave function
for the two channels.

The wave functions (\ref{chinc}) should be matched at $r \approx r_0$ to the
solution of the `inner' problem, i.e. that in the region of the strong
interaction. In the limit of small $r_0$ all the complexity of the `inner'
problem reduces to only two parameters. Namely, in the region of the strong
interaction the $n$ and $c$ channels are not independent and get mixed. Due to
the isotopic symmetry of the strong interaction the independent are the channels
with definite isospin, $I=0$ and $I=1$, corresponding to the functions
$\chi_0=\chi_n+  \chi_c$ and $\chi_1=\chi_n-  \chi_c$, and the matching
parameters are the logarithmic derivatives $-\kappa_0$ and $-\kappa_1$ of these
functions at $r=r_0$.  Using the assumption of small $r_0$ the matching
condition for the functions from Eq.(\ref{chinc}) can be shifted to $r =0$, so
that one can write the resulting matching equations as
\beq
{  k_n \cos \delta - \xi \kappa_c \over \sin \delta + \xi} = - \kappa_0~,~~~~~
{  k_n \cos \delta + \xi \kappa_c \over \sin \delta - \xi} = - \kappa_1~.
\label{match}
\eeq
These equations determine both the scattering phase $\delta$ and the constant
$\xi$ as
\beq
\cot \delta = -{\kappa_{\rm eff} \over k_n}
\label{cotdel}
\eeq
with
\beq
\kappa_{\rm eff}={2 \kappa_0 \kappa_1 -\kappa_c \kappa_1 - \kappa_c \kappa_0
\over
\kappa_0 + \kappa_1 - 2 \kappa_c}~,
\label{kapeff}
\eeq
and
\beq
\xi={\kappa_0 - \kappa_1 \over 2 \kappa_c - \kappa_1- \kappa_0} \, \sin \delta~.
\label{xires}
\eeq

The nonrelativistic scattering amplitude in the $n$ channel is therefore given
by\,\cite{ll}
\beq
f=- {1 \over \kappa_{\rm eff} + i \, k_n}~,
\label{ampf}
\eeq
and the scattering length $a$ is thus found from the $E =0$ limit of this
expression as
\beq
a=\left. {1 \over \kappa_{\rm eff}} \right |_{E=0}= {\kappa_0+\kappa_1- 2 \,
\sqrt{2
\mu \Delta} \over 2 \kappa_0 \kappa_1 - (\kappa_0 + \kappa_1) \sqrt{2 \mu
\Delta}}~.
\label{scata}
\eeq
The whole approach considered here is applicable if the scattering length is
large in the scale of strong interaction. A large positive value of $a$ implies
an existence of a shallow bound state, while a large negative $a$ corresponds to
the situation with a virtual state\,\cite{ll}. According to the estimates of
Ref.\cite{hkkn} the required by the data scattering length in the problem
considered is -(3 - 4)\,fm, corresponding to a negative and quite small indeed
parameter $\kappa_{\rm eff}(E=0)\approx$ (50 - 60)\,MeV.

The physical picture, consistent with a small $\kappa_{\rm eff}$, and which
could be
argued on general grounds\,\cite{ov} is that an attraction in the $I=0$ channel
is strong enough to provide a small value of $\kappa_0$, while the interaction
in the $I=1$ channel is either a weak attraction or, more likely, a repulsion.
In both cases the absolute value of $\kappa_1$ is large, i.e. of a normal strong
interaction scale, with the sign being respectively negative or positive.
Another, purely phenomenological, argument in favor of large $|\kappa_1|$ is
that no peculiar near-threshold behavior is observed in the production of the
$I=1$ charged states, e.g. $D^0 D^{*-}$. At large $|\kappa_1|$ the expression
(\ref{kapeff}) simplifies and takes the approximate form
\beq
\kappa_{\rm eff} \approx 2 \kappa_0 - \kappa_c~.
\label{kapefa}
\eeq
Using this approximation, one can readily see that in order for
$\kappa_{\rm eff}(E=0)$ to be negative and small the parameter $\kappa_0$ has to
be
positive and quite small:
\beq
\kappa_0 < \sqrt{\mu \Delta/2} \approx 63\,{\rm MeV}.
\label{ineq}
\eeq

It is interesting to note that in the suggested picture the interaction in the
$I=0$ state is strong enough by itself to produce a shallow bound state in the
limit of exact isospin symmetry, i.e. at $\Delta \to 0$. In reality the isospin
breaking by the mass difference between the charged and neutral charmed mesons
turns out to be sufficiently significant to deform the bound state into a
virtual one, i.e. to shift the pole of the scattering amplitude from the first
sheet to the second sheet of the Riemann surface for the amplitude as a complex
function of the energy $E$.

One can also notice that within the approximation in Eq.(\ref{kapefa}) the
scattering amplitude (\ref{ampf}) can be written in the form
\beq
f=-{1 \over 2 \kappa_0 - \kappa_c + i \, k_n}~,
\label{ampfa}
\eeq
which corresponds to equal coupling of the virtual state to the $n$ and $c$
channels. Such behavior, assumed in Ref.\cite{hkkn} on the grounds of isotopic
symmetry, turns out to be applicable, as long as $|\kappa_1|$ is large, even if
the isospin breaking by the mass difference is essential.

Thus far the presence of any inelastic channels was neglected in the discussion
of the $D {\bar D}^*$ scattering amplitude. Such channels certainly exist and
include the observed ones $\pi^+ \pi^- J/\psi$ ($\rho J/\psi$), $\pi^+ \pi^-
\pi^0 J/\psi$ ($\omega J/\psi$), $\gamma J/\psi$ and probably other, which are
yet to be found in experiment. The inelasticity however appears to be reasonably
small, as one can infer from the observed\,\cite{belledd,babardd} dominance of
the $D^0 {\bar D}^{*0}$ production in the threshold region, and can be
parametrized by a small imaginary shift $i \, \gamma$ of the denominator of the
scattering amplitude in Eq.(\ref{ampf}) or Eq.(\ref{ampfa}):
\beq
f=-{1 \over \kappa_{\rm eff}+ i \, k_n + i \, \gamma} \approx  -{1 \over  2
\kappa_0
- \kappa_c+ i \, k_n + i \, \gamma}~.
\label{ampfg}
\eeq
The latter expression is similar to the one used in Ref.\cite{hkkn} and differs
in that a term in the denominator, quadratic in $k_n$, being neglected, as
appropriate in the considered here small interaction radius approximation. The
previous analysis\,\cite{hkkn}, based on the Breit-Wigner type description,
includes such quadratic term and concludes that its contribution is very small
in the discussed region of parameters. Furthermore, the relatively small value
of the inelasticity is estimated there in terms of the scattering length as
Im$\,a$/Re$\,a \sim 0.1$ or less.

The width parameter $\gamma$ in Eq.(\ref{ampfg}) is the total sum over the
inelastic channels, and in what follows the cotributions of the $\omega J/\psi$
and $\rho J/\psi$ channels, $\gamma_\omega$ and $\gamma_\rho$, will be addressed
in some detail. It is further assumed here\,\cite{mv2,hkkn} that the `seed'
decay $B \to X K$ is a short-distance process, so that the yield in each final
channel coupled to $X$ is proportional to that channel's contribution to the
unitary cut of the amplitude $f$. This implies in particular that
\beq
{\cal B} [B \to (D^0 {\bar D}^{*0} + {\bar D} D^{*0}) \, K]\, :\, {\cal B} (B
\to \omega  J/\psi \, K)\, : \, {\cal B} (B \to \rho  J/\psi \, K) = \
k_n \, |f|^2 \, : \, \gamma_\omega \, |f|^2 \, : \, \gamma_\rho \, |f|^2~,
\label{ratio}
\eeq
where the specific expression for $|f|^2$ depends on the value of the energy $E$
relative to the $n$ and $c$ thresholds, as given by Eq.(\ref{ampfg}) an its
analytical continuation across the thresholds. Besides the energy dependence of
the overall factor $|f|^2$, the heavy meson channel contains the phase space
factor $k_n$, while for the $\omega J/\psi$ and $\rho J/\psi$ yields an
additional dependence on the energy arises from the factors $\gamma_\omega$ and
$\gamma_\rho$.

A certain variation of the width parameter $\gamma_\omega$ for the $\pi^+ \pi^-
\pi^0 J/\psi$ channel in the discussed range of energy is of a well known
kinematical origin. Indeed, the central value of the mass of the $\omega$
resonance puts the threshold for the channel $\omega J/\psi$ at
$3878.5\,$MeV, which corresponds to $E \approx 6.7\,$MeV in our conventions,
i.e. squarely between the $n$ and $c$ thresholds. Any production of the $\pi^+
\pi^- \pi^0 J/\psi$ states at smaller invariant mass is a sub-threshold process,
possible due to the width $\Gamma_\omega$ of the $\omega$ resonance. In other
words, the energy dependence of the width factor $\gamma_\omega$ can be
estimated as\cite{mv2,hkkn}
\beq
\gamma_\omega=|A_\omega|^2 \, q^{(\omega)}_{\rm eff}~,
\label{gw}
\eeq
where $A_\omega$ is the amplitude factor for the coupling to the $\omega J/\psi$
channel and $q^{(\omega)}_{\rm eff}$ is the effective momentum of $\omega$ at
the
invariant mass $M$ calculated as
\beq
q^{(\omega)}_{\rm eff} (M) = \int_{m_0}^{M-m_{J/\psi}} |{\vec q}(m)| \,
{m_\omega \,
\Gamma_\omega \over (m^2-m_\omega^2)^2 + m_\omega^2 \, \Gamma_\omega^2} \, {d
m^2 \over \pi}~
\label{qeff}
\eeq
with the c.m. momentum $|{\vec q}(m)|$ found in the standard way:
\beq
|{\vec q}(m)|={ \sqrt{ [(M-m_{J/\psi})^2- m^2] \, [(M+m_{J/\psi})^2- m^2]] }
\over 2 \, M}~.
\label{qofm}
\eeq
The lower limit $m_0$ in the integral in Eq.(\ref{qeff}) can be chosen anywhere
sufficiently below $m_\omega - \Gamma_\omega$, since the Breit-Wigner curve in
the integrand rapidly falls off away from the resonance.

Numerically, the effective momentum $q^{(\omega)}_{\rm eff}$ can be estimated as
varying from approximately 20\,MeV to 50\,MeV between the $n$ and $c$
thresholds, i.e. when $E$ changes from $E=0$ to $E = \Delta$. As will be argued
here, the amplitude $A_\omega$ should vary only slowly in the considered energy
range, so that the shape of the threshold cusp in the $\omega J/\psi$ channel is
determined by the behavior of the scattering amplitude factor $|f|^2$ and by the
estimated kinematical effect.

In the $\rho J/\psi$ channel the expected energy behavior of the yield is quite
different. If one writes the corresponding width factor $\gamma_\rho$ similarly
to Eq.(\ref{gw}) as
\beq
\gamma_\rho= |A_\rho|^2 \, q^{(\rho)}_{\rm eff}~,
\label{gr}
\eeq
the effective momentum $q^{(\rho)}_{\rm eff}$ can be estimated a s varying only
slightly due to the large width of the $\rho$ resonance: $q^{(\rho)}_{\rm eff}
\approx (125 \div 135)\,$MeV as the energy changes between $E=0$ and $E=\Delta$.
On the contrary, as will be argued, the amplitude $A_\rho$ should experience a
significant variation in this energy range  and in fact cross zero between the
$n$ and $c$ thresholds.

In order to argue the claimed properties of the amplitudes $A_\omega$ and
$A_\rho$ it can be first noticed that the inelastic processes $D {\bar D}^* \to
\omega J/\psi$ and $D {\bar D}^* \to \rho J/\psi$ involve a rearrangement of the
heavy and light quarks and therefore cannot be due to peripheral interactions at
long distances, but are determined by the dynamics at a typical range of the
strong interaction. In other words the amplitudes of these processes are
sensitive to the behavior of the heavy meson pair wave function at short
strong-interaction distances. Thus for the $I=0$ state $\omega J/\psi$ the
amplitude is related to the short-distance part of the function $\chi_0$, while
the amplitude for the $I=1$ channel $\rho J/\psi$ is determined by the function
$\chi_1$. Assuming also as previously, that the $B$ decays are also determined
by short distances, one can write for the decay amplitudes the expressions
\beq
A(B \to \omega J/\psi \,K) = \int F_\omega \, \chi_0(r) \, dr~~~{\rm and}~~~A(B
\to \rho J/\psi \,K) = \int F_\rho \, \chi_1(r) \, dr~,
\label{overlap}
\eeq
where $F_\omega$ and $F_\rho$ are the weight functions in the respective
channels, and their support, as argued, is limited to typical
strong-interaction distances\footnote{Certainly, at present any details of these
functions are not known. It can only be mentioned that it is likely that $F_\rho
\approx F_\omega$ due to the $\rho -\omega$ universality.}. In the limit of
vanishing interaction range these functions should each be replaced by a
$\delta$-function: $F_{\omega,\rho} \to \Phi_{\omega,\rho} \, \delta(r)$, so
that, using the expressions in Eq.(\ref{chinc}) and the results in the equations
(\ref{cotdel}), (\ref{kapeff}) and (\ref{xires}) for the solution to the
matching conditions, one finds
\begin{eqnarray}
&&A(B \to \omega J/\psi \,K) = 2 \, {\kappa_1- \kappa_c \over
\kappa_0+\kappa_1-2 \, \kappa_c} \,  \Phi_\omega \,  \sin \delta  \approx 2 \,
\Phi_\omega \,   \sin \delta  \nonumber \\
&& A(B \to \rho J/\psi \,K) = 2 \, {\kappa_0- \kappa_c \over \kappa_0+\kappa_1-2
\, \kappa_c} \,  \Phi_\rho \,  \sin \delta  \approx 2\, {\kappa_0- \kappa_c
\over \kappa_1}\, \Phi_\rho \,   \sin \delta~,
\label{pointa}
\end{eqnarray}
where the latter expressions for each amplitude are written in the limit of
large $|\kappa_1|$. One can readily see that in this limit the amplitude for
production of the isoscalar state $\omega J/\psi$ is finite, while that for the
isovector state $\rho J/\psi$ is suppressed by $1/\kappa_1$. Furthermore, it
should be noted that, generally, in the limit of small interaction range it
would be incorrect to retain such suppressed term without taking also into
account effects of finite range. In order to fix this deficiency in the second
line in Eq.(\ref{pointa}) one should go beyond the $r=0$ point approximation for
the second integral in Eq.(\ref{overlap}) and use the first two terms of the
Taylor expansion of the function $\chi_1(r)$ at $r=0$ rather than only the first
term:
\beq
\chi_1(r)=\sin \delta - \xi + (k_n \, \cos \delta + \xi \,
\kappa_c)\, r + O(r^2) =  2 \, {\kappa_0 - \kappa_c \over
\kappa_0+\kappa_1-2 \, \kappa_c} \, ( 1 -  \, \kappa_1 \, r) \, \sin
\delta + O(r^2)~.
\label{taylor}
\eeq
Then the improved estimate of the second integral in Eq.(\ref{overlap}) can be
written in terms of the effective radius $R$ of the weight function $F_\rho$,
\beq
R= {\int F_\rho \, r \, dr \over \int F_\rho \, dr}
\label{reff}
\eeq
as
\beq
A(B \to \rho J/\psi \,K) = 2 \, {\kappa_0- \kappa_c \over
\kappa_0+\kappa_1-2 \, \kappa_c} \, (1 -  \, \kappa_1 \, R)\,
\Phi_\rho \,  \sin \delta  \approx 2 \,(\kappa_0- \kappa_c) \left
({1 \over \kappa_1}-R \right )\, \Phi_\rho \, \sin \delta~,
\label{improv}
\eeq
where, as discussed, the parameter $\kappa_1 R$ should be considered as being of
order one.

Clearly, the factor $\sin \delta$ in each amplitude is proportional to the
scattering amplitude factor $f$, which is accounted for separately in the yield
of each of the discussed channels (Eq.(\ref{ratio})). Therefore this factor
should be omitted in the amplitudes $A_\omega$ and $A_\rho$, and one
can conclude that the amplitude $A_\omega$ is a smooth slowly varying function
of the energy inasmuch as $\kappa_1$ is large. Furthermore, the ratio of the
$\rho J/\psi$ and $\omega J/\psi$ production amplitudes is estimated as
\beq
{A_\rho \over A_\omega} = {\kappa_0-\kappa_c \over \kappa_1-\kappa_c} \, (1
-\kappa_1 \, R) \, {\Phi_\rho \over \Phi_\omega}~.
\label{rorat}
\eeq
Since the situation where the $X$ peak is a virtual state corresponds to a small
positive $\kappa_0$ satisfying the condition (\ref{ineq}), the amplitude
$A_\rho$ described by Eq.(\ref{rorat}) should necessarily change sign between
the $n$ threshold, where $\kappa_c=\sqrt{2 \mu \Delta}$, and the $c$ threshold,
where $\kappa_c=0$. It can be noticed that the existence of the zero of the
amplitude $A_\rho$ results from both first terms of the expansion (\ref{taylor})
for the function $\chi_1(r)$ being proportional to $\kappa_0 - \kappa_c$.

\begin{figure}[ht]
  \begin{center}
    \leavevmode
    \epsfxsize=10cm
    \epsfbox{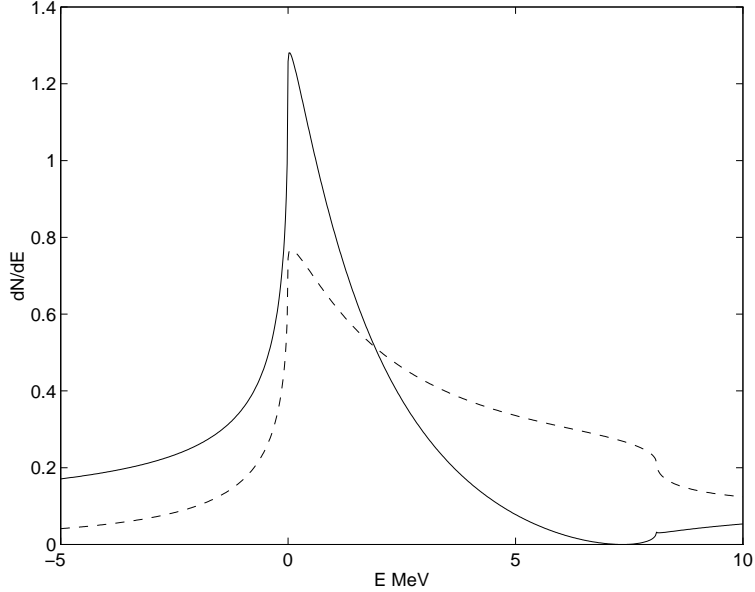}
    \caption{The expected shape (in arbitrary units) of the virtual state peak
in the yield of $\pi^+ \pi^- J/\psi$ (solid) and $\pi^+ \pi^- \pi^0 J/\psi$
(dashed) channels.}
  \end{center}
\end{figure}

The expected difference in the shape of the cusp in the $\rho J/\psi$ and
$\omega J/\psi$ channels is illustrated in Fig.1. In these plots the parameters
of the virtual state correspond to the scattering length $a=-(4 + 0.5\, i)\,$fm,
which is close to the possible fit values of the scattering length found in
Ref.\cite{hkkn}. In the limit of large $\kappa_1$ this value of $a$ translates
into $\kappa_0 \approx 38\,$MeV and $\gamma \approx 6\,$MeV. One can see from
Fig.1 that due to the discussed zero of the amplitude the peak in the $\pi^+
\pi^- J/\psi$ is expected to be quite narrow in agreement with the experimental
limit on the width of $X(3872)$. The plots in Fig.1 are normalized to the same
total yield in each channel over the shown energy range in order to approximate
the experimentally observed relative yield. Such normalization
corresponds to setting
$$\left | {\kappa_1 \over 1-\kappa_1 R} \, {\Phi_\omega \over \Phi_\rho} \right|
\approx
175\,{\rm MeV}~,$$
which value does not appear to be abnormal, even though at present we have no
means of independently estimating this quantity.

Clearly, the suggested significant difference of the shape of the virtual state
peak in the two discussed channels provides with a way of discriminating between
the options of $X$ being a virtual state and a bound state.

I acknowledge enlightening discussions of general properties of a low-energy
scattering with Arkady Vainshtein and helpful communications with
Yulia Kalashnikova and Alexander Kudryavtsev regarding their paper \cite{hkkn}.
This work is supported in part by the DOE grant DE-FG02-94ER40823.

\end{document}